\documentclass{article}
\usepackage{epsfig,natbib,emulateapj,apjfonts}

\newcommand{\msun}{\hbox {$M_{\sun}$ }}

\newcommand{\logg}{\hbox {$\log g\,$}}       
\begin{document}

\submitted{to appear in ApJL}

\title{The Pulsation Properties of Procyon A}
\author{Brian Chaboyer}
\affil{Department of Physics and Astronomy, Dartmouth College
6127 Wilder Laboratory, Hanover, NH, USA 03755-3528}

\author{P.\ Demarque }
\affil{Department of Astronomy, and Center for Solar and
Space Research, Yale University, Box 208101, New Haven, CT 06520-8101}

\and

\author{D.B.\  Guenther}
\affil{Department of Astronomy and Physics, Saint Mary's University, 
Halifax, Nova Scotia, Canada B3H 3C3 }

\begin{abstract}
A grid of stellar evolution models for Procyon A has been calculated.
These models include the best physics available to us (including the
latest opacities and equation of state) and are based on the revised
astrometric mass of \citet{mass}.  Models were calculated with helium
diffusion and with the combined effects of helium and heavy element
diffusion.  Oscillation frequencies for $\ell=0,1,2$ and 3 $p$-modes
and the characteristic period spacing for the $g$-modes were
calculated for these models.  We find that $g$-modes are sensitive to
model parameters which effect the structure of the core, 
such as convective core overshoot, the heavy element abundance and the
evolutionary state (main sequence or shell hydrogen burning) of
Procyon A.. The $p$-modes are relatively insensitive to the
details of the physics used to model Procyon A, and only depend on the
evolutionary state of Procyon A.  Hence, observations of
$p$-mode frequencies on Procyon A will serve as a robust test of
stellar evolution models.

\end{abstract}

\keywords{binaries: visual --- stars: evolution --- stars: 
individual (Procyon, alpha Can Min) --- stars: oscillations}

\section{Introduction}
Procyon's internal structure makes it a particularly interesting
target for seismological studies.  Because models of
stars at Procyon's position in the HR-diagram have convective
envelopes and because we know for the Sun that the solar convection
zone is responsible for driving its $p$-mode oscillations, we do
expect to see $p$-mode oscillations on Procyon. Furthermore, because
Procyon's convective envelope is very thin, any $g$-modes that exist,
possibly driven by Procyon's convective core, may be observable at the
surface, not being completely damped out as they are in the case of
the Sun with its more extensive convective envelope.

Seismology would establish the evolutionary status of Procyon A, which
could be either in the core hydrogen burning phase, or in the more
advanced hydrogen shell burning phase, after the core has been
exhausted and is composed purely of helium.  In the latter case, the
presence of an exhausted core and a mean molecular weight
discontinuity would leave an unmistakable signature on the $p$-mode
frequency spacings due to the existence of mode
bumping where the $g$-mode spectrum overlaps (bumps into) and perturbs
the $p$-mode spectrum. Mode bumping affects the spacing between
adjacent $p$-mode frequencies, and hence, can be discerned from non-bumped
$p$-modes. The $g$-mode spectrum would, of course, provide even more
direct clues as to the exact nature of Procyon's core

The Procyon binary system consists of an F5 IV-V primary and a
white-dwarf secondary in an 40.8 year orbit.  The F5 primary (Procyon
A) is a bright, nearby star with a well determined parallax and
astrometric mass.  As such, it presents a unique target for the study
of non-radial stellar oscillations.  Because of the large difference
in brightness between the two stars in Procyon, the orbital parameters
have been difficult to measure.  \citet{strand} derived a mass of
$1.75\, \msun$ for Procyon A, a mass which was incompatible with
astrophysical estimates from stellar interior models \citep{gd93} and
the spectroscopic \logg.  Only recently, with improved orbital
elements and parallax, and a precise angular scale derived using the
cold chronograph (CoCo) at the NASA IR Telescope Facility (IRTF)
\citep{wang}, has it been possible to redetermine the mass of Procyon A
\citep{mass}.  The revised mass, about $1.5\, \msun$, is now in
agreement with the astrophysical mass estimate.

\citet{paperI}, stimulated by the SONG ground-based campaign
\citep{song} to detect
oscillations on Procyon, made a study of the evolutionary
status of Procyon A, using up-to-date physical input and the revised
Procyon A mass.  Because the SONG data analysis is not yet
complete, only a brief summary of these Procyon models was published

However, the situation has changed dramatically in the last few weeks,
with the successful identification of individual mode frequencies in
the giant star $\alpha$ UMa A by \citet{wire}, using the small
telescope on the WIRE space mission. Because Procyon is on the WIRE
target list for observations in the near future, theoretical models
and their predicted frequencies have become particularly relevant.  We
present in this paper a more detailed description of our interior
models, and a list of individual $p$-mode frequencies for our reference
model.  We also include a table of averaged large and small $p$-mode
spacings and the characteristic $g$-mode period spacings for our
models.
\begin{deluxetable}{llcllllccccccl} 
\tablefontsize{\scriptsize}
\tablecaption{Model Characteristics \label{tab1}}
\tablewidth{0pt}

\tablehead{
&\colhead{Mass} &
\colhead{$\,\,\Delta Y$} & 
&
\colhead{${\rm M_{core}}$}&
\colhead{${\rm M_{scz}}$}&
\colhead{$\log$}&
\colhead{$<\!\!\Delta\nu_0\!\!>$}&
\colhead{$<\!\!\Delta\nu_1\!\!>$}&
\colhead{$<\!\!\Delta\nu_2\!\!>$}&
\colhead{$<\!\!\Delta\nu_3\!\!>$}&
\colhead{$<\!\!\delta\nu_0\!\!>$}&
\colhead{$<\!\!\delta\nu_1\!\!>$}&
\colhead{$\Pi$}\\
\colhead{Model}&
\colhead{(${\rm M}_{\odot}$)}&
\colhead{$\overline{\Delta Z}$}&
\colhead{$Z/X_{\rm env}$}&
\colhead{(${\rm M}_{\odot}$)}&
\colhead{(${\rm M}_{\odot}$)}&
\colhead{$({\rm L/L}_{\odot})$}&
\colhead{($\mu$Hz)}&
\colhead{($\mu$Hz)}&
\colhead{($\mu$Hz)}&
\colhead{($\mu$Hz)}&
\colhead{($\mu$Hz)}&
\colhead{($\mu$Hz)}&
\colhead{($\mu$Hz)}
}
\startdata

Standard & 1.50 &   3.03  & 0.0253 & 0.118 & 
1.17E--4 & 0.85868  & 53.95 & 54.15 &  51.38 &  54.40 & 4.75  &    7.01&
60.58\\

subgiant & 1.40 &   2.94  & 0.0239 &  0.000 & 
5.62E--5 & 0.85838 & 51.89 & 48.71 &  46.23 &  45.64 &  6.03  & 7.35 &
16.87 \\

low L    & 1.50 &  2.47  & 0.0243 &  0.113 & 
2.31E--4 & 0.83747 & 54.19 &  54.34&  51.62&  51.80&     4.69 &    7.06&
 59.35 \\

no wind  & 1.50 &  3.12  & 0.0196 &  0.122 & 
1.36E--5 & 0.85852 & 52.99 & 53.22&  50.50&  50.69&     5.09&     7.26&
62.02 \\

overshoot& 1.50 &  2.53  & 0.0248 &  0.133 & 
1.04E--4 & 0.85848 & 53.80 &54.06&  51.28&  54.25&     4.64&     6.92&
64.77 \\

low Z    & 1.50 &  2.02   & 0.0206 &  0.107 & 
1.15E--4 & 0.85860 & 53.87 & 54.05&   51.41&  54.52&     5.23&
7.10 & 56.35\\

high M   & 1.54 &  1.90  & 0.0247 &  0.121 &
1.07E--4 & 0.85854 & 54.75 & 54.95 & 54.97 & 52.32 &    4.58 &    7.17&
61.65 \\

Z diff   & 1.50 &  3.01  & 0.0253 & 0.118 & 
1.19E--4 & 0.85847 & 53.94 & 54.15&  51.38&  54.40 &    4.73&     7.01&
60.70 \\

\enddata
\end{deluxetable}

\vspace*{1.0cm}

\section{Modeling Procyon A}
Models for Procyon A were calculated using the Yale stellar evolution
code, in its non-rotating configuration \citep{yrec}.  These models
included the latest OPAL opacities \citep{opal}, and the OPAL equation
of state \citep{opaleos}. The diffusion coefficients are from
\citet{thoul}.  Most of the models only included helium diffusion.  A
single model which included both helium and heavy element diffusion
(treated as a single mean heavy element, $Z$) was evolved to study the
effects of heavy element diffusion on Procyon A.  The models included
the effects of a wind mass loss in the diffusion equations.  A wind
velocity of $v_w = -\dot{\rm M}/(4\pi \rho r^2)$ was included in the
diffusion equations, and a solar mass loss rate $\dot{\rm M} = 2\times
10^{-14}\,\msun/{\rm yr}$ was assumed.  This wind velocity was large
enough to effectively suppress the diffusion in the outer layers of
the model.  One model was calculated without wind loss.  The models
employed a solar calibrated mixing length, though the main sequence
models of Procyon A have a very thin surface convection zone and so
the structure of these models are largely independent of the mixing
length.  None of our models includ the effects of rotation. As Procyon
A is not a rapidly rotating star, the structure effects of rotation
will be much smaller than the other factors considered in this paper.
Rotation likely leads to mixing in stars, and this mixing can supress
diffusion in the outer layers of the star \citep{rotdiff}. This effect
is similar to the inclusion of a stellar wind (which also suppresses
diffusion in the outer layers), and so a rough estimate of the
importance of rotational mixing can be made by comparing the standard
case to the no wind case.

The fundamental properties of Procyon A are: $M = 1.50\pm 0.05\,\msun$
\citep{mass}, $\pi = 0.2832''\pm 0.0015$ \citep{mass}, $f_\oplus = (18.64\pm
0.87)\times 10^{-6}\, {\rm ergs\, cm^{-2}}\, s{^-1}$ \citep{small95}
and an angular diameter of $\phi = 5.51\pm 0.05\,$mas \citep{moz91}
which imply $L= (7.22\pm 0.35)\,{\rm L}_{\odot}$ and $R = (2.09\pm
0.02)\,{\rm R}_{\odot}$.  The central values for the mass, radius and
luminosity were adopted for the standard model of Procyon A (see
below). The models for Procyon A were evolved from the ZAMS until they reached
the observed radius.  In an iterative procedure, the helium abundance
was adjusted in the ZAMS model and the model was re-evolved, until the
model matched the observed radius and luminosity of Procyon A.  The
starting helium abundance was constrained such that $1.5 < \Delta
Y/\Delta Z < 4$. In addition, the final surface abundance of $Z/X$ was
constrained to be within 0.1 dex of the solar value $Z/X = 0.0245$
\citep{gre93}, to satisfy the observed constraint that the surface
abundances of Procyon A are near solar
\citep{takeda94,takeda96,kato96,yush96}.  Models which met all of the
above criteria had their pulsation frequencies calculated using
Guenther's non-radial non-adiabatic stellar pulsation program
\citep{guenth94}.

In total, eight different calibrated models of Procyon A were
calculated and pulsed (Table \ref{tab1}).  The standard model (line 1
in Table \ref{tab1}) uses the best estimate for the mass, radius and
luminosity, does not include heavy element diffusion or overshoot at
the convective boundaries, and uses our calibrated solar value for the
mass fraction of the heavy elements ($Z = 0.018$).  The other seven
models involved changing a single parameter from the standard model:
(a) lower mass (subgiant); (b) evolution to a somewhat lower
luminosity (low L); (c) no wind loss; (d) convective core overshoot of
0.1 pressure scale heights (overshoot); (e) low heavy element
abundance (low $Z$, with $Z = 0.015$); (f) higher mass (high M); and
(g) heavy element diffusion ($Z$ diff).  All but the subgiant model
are in the main sequence phase of evolution.  The lower mass model is
in the hydrogen shell burning phase of evolution.

\section{Pulsation Properties}  
The average first-order frequency spacing $<\!\Delta\nu\!>$ of the
$p$-modes, with $\Delta \nu(n,\ell) \equiv \nu(n,\ell) -
\nu(n-1,\ell)$, is likely to be the first quantity determined by
stellar seismology.  Table \ref{tab1} includes the frequency spacing
averaged over $n = 5$ to 25  for $\ell = 0,\,1,\,2$ and
$\ell = 3$.  The first-order frequency spacing is sensitive to the
structure of the outermost layers of the star.  Table
\ref{tab1} also includes the average second-order frequency spacing
($<\!\delta\nu\!>$, averaged over $n = 5$ to 25) for $\ell = 0$ and
1.  The second order frequency spacings, $\delta \nu(n,\ell) \equiv
\nu(n,\ell) - \nu(n-1,\ell + 2)$, are sensitive to the structure of
the innermost layers of the star.  Table \ref{tab1}
includes the characteristic period spacing $\Pi$ of the $g$-modes.

A list of modes from $\ell = 0$ to 3 and $n = 0$ to 24 for the
standard model is given in Table \ref{tab2}.  In this table, $n_p$ is
the number of nodes in the eigenfunction that are $p$-mode like (with
respect to their phase characteristics), and $n_g$ is the number of
nodes in the eigenfunction that are $g$-mode like. The gaps in the
spacings are due to the $n$-value naming convention \citep{unno89}.
In regions of variable mean molecular weight, the frequencies of the
$g$-modes trapped in the interior may overlap with the frequencies of
the $p$-modes trapped in the envelope, "bumping" the $p$-modes, and
resulting in the mixing of $p$-mode like modes and $g$-mode like modes
in the $p$-mode eigenfunction.
As a consequence of mode 
bumping the large spacing varies irregularly in the frequency range 
where mixed modes (i.e.\ mode bumping) occurs. In the standard  
model only the lowest frequency $p$-modes (lowest $n$) are mixed but for the 
subgiant model, all of the nonradial $p$-modes we calculated, up to $n = 
25$, are mixed. 
\centerline{\epsfig{file=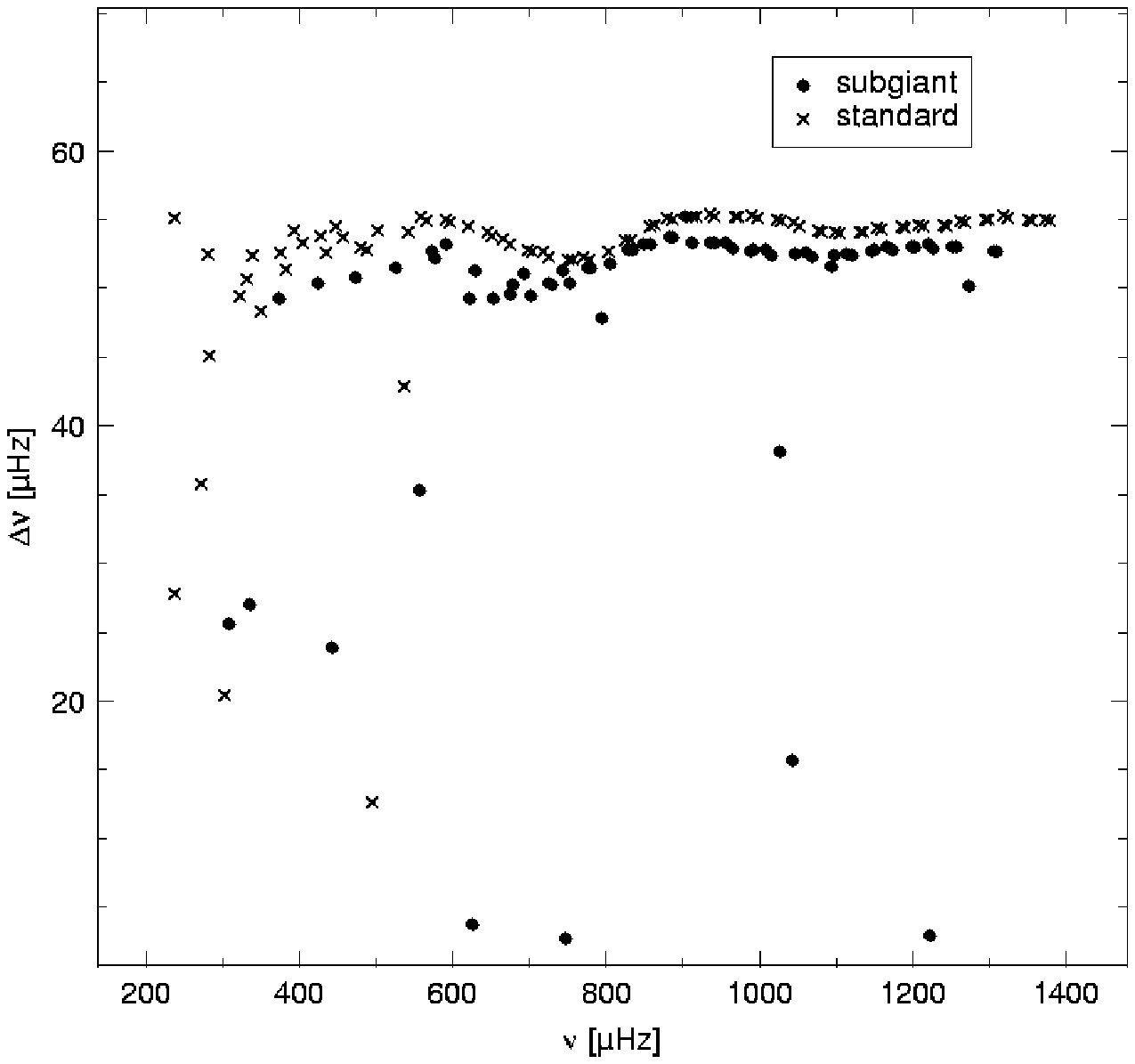,height=8.0cm}}
{Fig.\ 1.The large frequency spacing $\Delta\nu$ is
plotted as a function of frequency $\nu$ for the standard and
subgiant models.  Note that irregularities in the large frequency
spacings extend to higher frequencies for the subgiant model as
compared to the standard model. }
\vspace*{0.5cm}

\noindent
In Figure 1 we show the large spacing for the standard model and for the 
subgiant model. As just described, we see that the irregularities in the 
large spacings extend to higher frequencies for the subgiant model than 
the standard model.

The frequencies and characteristic frequency spacings depend on the
radius of the star.  To first order, $\delta {\rm R_*/R_*} \simeq
\delta \nu/\nu$.  The radius of Procyon A is known to within 1\% .
Thus, the calculated pulsation frequencies are uncertain at the 1\%
level due to the error in the radius.  In addition, uncertainties in
the modeling of the superadiabatic layer (in both the evolution and
pulsation calculations) leads to an estimated uncertainty of $\sim \pm
0.5\%$ in the $p$-mode frequency calculations (see \cite{gd96}).  The
total error associated with the calculated pulsation frequencies shown
in Table \ref{tab1} is approximately $\sim 1.5\%$.  Thus, differences
in $\Delta\nu$ greater than $\sim 1\mu$Hz are significant.  From Table
\ref{tab1} we see that only the subgiant model has  $\Delta\nu$ which are
significantly different from the others.  When the large spacing is
observed on Procyon, it will immediately tell us whether or not
Procyon has reached the subgiant phase of evolution.

The characteristic period spacing ($\Pi$) of the $g$-modes shows a
much larger variation than $\Delta\nu$ (Table \ref{tab1}).  Once again,
due to errors in the models and radii of Procyon A, only differences
in $\Pi$ greater than $\sim 1\mu$Hz are significant.  In this context,
it is clear that the detection of $g$-modes in Procyon A would allow
one to (a) determine its evolutionary status; and (b) determine if
appreciable overshoot ($\ga 0.05\,{\rm H_p}$) is occurring at the edge
of the convective core in Procyon A, and/or provide an estimate of the
interior metallicity of Procyon A.

\section{Summary}
A grid of stellar evolution models for Procyon A has been calculated
and their pulsation properties investigated. The grid of models was
chosen to examine the effect of varying the physics used to construct
the models (such as the inclusion of overshoot, heavy element
diffusion or a stellar wind) as well as uncertainties in the observed
parameters of Procyon A (mass, luminosity, radius and heavy element
composition).  The properties of these models (including the average
frequency spacing for $p$ and $g$-modes) are listed in Table
\ref{tab1}, while Table \ref{tab2} presents a list of individual
frequencies for our standard model of Procyon A.  The predicted
$p$-mode frequencies are fairly similar for all models, except for the
model which uses a lower ($1.4\,\msun$) mass for Procyon A.  This
model is in the subgiant (shell hydrogen burning) phase of
evolution, while all other models are on the main sequence.  If
Procyon A is a main sequence star, the predicted frequency spacings
are fairly regular, while mode bumping leads to irregular frequency
spacings in the subgiant star.  As the predicted frequencies are
relatively insensitive to changes in the input physics, or the assumed
mass, radius, luminosity and heavy element abundance of Procyon, the
detection of $p$-modes on Procyon A will serve as a robust test of
stellar evolution theory.  Procyon A is on the WIRE target list for
observations in the near future.  As WIRE has been successful in
detecting individual mode frequencies in the giant star $\alpha$ 
UMa A \citep{wire}, it is hoped that the predictions presented here
for the pulsation properties of Procyon A will soon be tested.  

\acknowledgements
This work was supported in part by an Canadian NSERC grant to D.B.G.

\begin{deluxetable}{lrrlrllllrrlrl}
\tablecolumns{14}
\tablewidth{0pt}
\tablenum{2}
\tablecaption{Standard Model Frequencies (in $\mu$Hz)\label{tab2}}
\tablehead{
\colhead{$\ell$}&
\colhead{$n$}&
\colhead{$n_p$}&
\colhead{$n_g$}&
\colhead{$\nu$}&
\colhead{$\Delta\nu$}&
\colhead{$\delta\nu$}&&
\colhead{$\ell$}&
\colhead{$n$}&
\colhead{$n_p$}&
\colhead{$n_g$}&
\colhead{$\nu$}&
\colhead{$\Delta\nu$}
}
\startdata
 0&  0&   0&   0&  129.02& \nodata&    \nodata &&  2&  0&   2&   2&  208.98&  \nodata \\
 0&  2&   2&   0&  227.83& \nodata&    \nodata &&  2&  1&   2&   1&  236.75&  27.77   \\        
 0&  3&   3&   0&  280.35&  52.52 &    7.84    &&  2&  2&   3&   1&  272.51&  35.76   \\
 0&  4&   4&   0&  330.93&  50.59 &    8.96    &&  2&  3&   4&   1&  321.97&  49.47   \\
 0&  5&   5&   0&  382.26&  51.32 &    7.71    &&  2&  4&   5&   1&  374.55&  52.57   \\
 0&  6&   6&   0&  434.86&  52.61 &    6.47    &&  2&  5&   6&   1&  428.39&  53.84   \\
 0&  7&   7&   0&  487.66&  52.80 &    6.26    &&  2&  6&   7&   1&  481.40&  53.01   \\
 0&  8&   8&   0&  541.74&  54.08 &    \nodata &&  2&  7&   7&   0&  494.03&  12.63   \\ 
 0&  9&   9&   0&  596.51&  54.76 &    \nodata &&  2&  8&   8&   0&  536.95&  42.92   \\
 0& 10&  10&   0&  650.31&  53.80 &    \nodata &&  2&  9&   9&   0&  591.97&  55.01   \\ 
 0& 11&  11&   0&  702.97&  52.66 &    \nodata &&  2& 10&  10&   0&  646.02&  54.06   \\ 
 0& 12&  12&   0&  755.02&  52.05 &    \nodata &&  2& 11&  11&   0&  698.82&  52.80   \\ 
 0& 14&  13&   0&  807.81&  \nodata&   4.21    &&  2& 12&  12&   0&  750.94&  52.11   \\
 0& 15&  14&   0&  862.40&  54.60 &    4.33    &&  2& 13&  13&   0&  803.60&  52.66   \\
 0& 16&  15&   0&  917.64&  55.23 &    4.35    &&  2& 14&  14&   0&  858.07&  54.47   \\
 0& 17&  16&   0&  972.79&  55.16 &    4.35    &&  2& 15&  15&   0&  913.28&  55.21   \\
 0& 18&  17&   0& 1027.70&  54.91 &    4.28    &&  2& 16&  16&   0&  968.45&  55.16   \\
 0& 19&  18&   0& 1081.82&  54.12 &    4.21    &&  2& 17&  17&   0& 1023.42&  54.97   \\
 0& 20&  19&   0& 1135.90&  54.08 &    4.21    &&  2& 18&  18&   0& 1077.61&  54.19   \\
 0& 21&  20&   0& 1190.29&  54.39 &    4.16    &&  2& 19&  19&   0& 1131.70&  54.08   \\
 0& 22&  21&   0& 1244.81&  54.51 &    4.10    &&  2& 20&  20&   0& 1186.13&  54.44   \\
 0& 23&  22&   0& 1299.77&  54.97 &    4.02    &&  2& 21&  21&   0& 1240.71&  54.58  \\
 0& 24&  23&   0& 1354.69&  54.91 &    3.87    &&  2& 22&  22&   0& 1295.76&  55.05  \\
&\nodata&\nodata&\nodata&\nodata&\nodata&\nodata&&  2& 23&  23&   0& 1350.82&  55.06  \\[10pt]
 1& -2&   0&   2&  129.55&  \nodata&  \nodata &&  3&  2&   3&   1&  286.72&  \nodata  \\
 1&  0&   1&   1&  181.62&  \nodata&  \nodata &&  3&  3&   4&   1&  339.15&  52.43    \\
 1&  1&   2&   1&  236.71&  55.08  &  \nodata &&  3&  4&   5&   1&  393.34&  54.19    \\
 1&  2&   3&   1&  281.78&  45.07  &  \nodata &&  3&  5&   6&   1&  447.87&  54.53    \\ 
 1&  3&   4&   1&  302.21&  20.43  &  15.49   &&  3&  6&   7&   1&  502.07&  54.20    \\
 1&  4&   5&   1&  350.49&  48.28  &  11.34   &&  3&  7&   8&   1&  557.30&  55.23    \\
 1&  5&   6&   1&  403.82&  53.32  &  10.47   &&  3&  9&   9&   0&  612.62&  \nodata  \\
 1&  6&   7&   1&  457.48&  53.66  &   9.61   &&  3& 10&  10&   0&  666.24&  53.62    \\
 1&  8&   9&   1&  511.24&  \nodata&  \nodata &&  3& 11&  11&   0&  718.88&  52.65    \\
 1&  9&  10&   1&  566.13&  54.89  &  \nodata &&  3& 12&  12&   0&  771.10&  52.22    \\
 1& 10&  10&   0&  620.65&  54.52  &   8.03   &&  3& 13&  13&   0&  824.58&  53.47    \\
 1& 11&  11&   0&  673.85&  53.20  &   7.62   &&  3& 14&  14&   0&  879.65&  55.08    \\
 1& 12&  12&   0&  726.14&  52.29  &   7.26   &&  3& 15&  15&   0&  935.03&  55.37    \\
 1& 13&  13&   0&  778.19&  52.04  &   7.09   &&  3& 16&  16&   0&  990.31&  55.29    \\
 1& 14&  14&   0&  831.69&  53.51  &   7.12   &&  3& 17&  17&   0& 1045.11&  54.80    \\
 1& 15&  15&   0&  886.65&  54.96  &   7.00   &&  3& 18&  18&   0& 1099.24&  54.13    \\
 1& 16&  16&   0&  941.82&  55.16  &   6.79   &&  3& 19&  19&   0& 1153.64&  54.40    \\
 1& 17&  17&   0&  996.92&  55.10  &   6.61   &&  3& 20&  20&   0& 1208.28&  54.64    \\
 1& 18&  18&   0& 1051.45&  54.53  &   6.34   &&  3& 21&  21&   0& 1263.15&  54.87    \\
 1& 19&  19&   0& 1105.40&  53.95  &   6.17   &&  3& 22&  22&   0& 1318.42&  55.27    \\
 1& 20&  20&   0& 1159.70&  54.30  &   6.06   &&  3& 23&  23&   0& 1373.45&  55.03    \\
 1& 21&  21&   0& 1214.21&  54.51  &   5.93   \\
 1& 22&  22&   0& 1269.01&  54.80  &   5.86   \\
 1& 23&  23&   0& 1324.15&  55.14  &   5.72   \\
 1& 24&  24&   0& 1378.99&  54.85  &   5.54   \\

\enddata 
\end{deluxetable} 
\end{document}